\documentclass[11pt]{article}
\usepackage[a4paper,margin=2.5cm]{geometry}
\usepackage[colorlinks=true,allcolors=blue]{hyperref}

\usepackage[utf8]{inputenc}
\usepackage{graphicx}
\usepackage{amsmath}
\usepackage{amssymb}

\begin{document}

\title{Coherent phase slips in coupled rings: from weak to strong interactions}

\author{A. Perez-Obiol$^{1,*}$, M.G. Boshier$^{2}$, L. Amico$^{3,4,5}$ and J. Polo$^{3}$\\[2ex]
\small $^1$Departament de Física, Universitat Autònoma de Barcelona, 08193, Bellaterra, Spain\\
\small $^2$Materials Physics and Applications Division, Los Alamos National Laboratory,\\
\small Los Alamos, New Mexico 87545, USA\\
\small $^3$Quantum Research Centre, Technology Innovation Institute, Abu Dhabi, UAE\\
\small $^4$Dipartimento di Fisica e Astronomia, Via S. Sofia 64, 95127 Catania, Italy\\
\small $^5$INFN-Sezione di Catania, Via S. Sofia 64, 95127 Catania, Italy\\[1ex]
\small $^*$E-mail: \href{mailto:axel.perezobiol@uab.cat}{axel.perezobiol@uab.cat}}

\date{}
\maketitle

\begin{abstract}
We study the coherent transfer of circulation between two weakly coupled lattice rings of ultracold bosons, in which a persistent current initially prepared in one ring oscillates back and forth between the rings without net transfer of particles. Combining Bose--Hubbard simulations with their Gross--Pitaevskii and quantum phase model limits, we show that similar current oscillations arise from two distinct microscopic mechanisms. At weak interactions, particles in each occupied momentum mode tunnel independently between the rings, leading to a transfer time inversely proportional to the inter-ring tunnelling, extensive particle-number fluctuations and entanglement distributed over many Schmidt components; this mode-exchange dynamics is suppressed by self-trapping as the interactions increase. At stronger interactions and large filling, in the phase-dominated regime, circulation is instead transferred through a second-order collective process between many-body persistent-current states, with a transfer time scaling as the inverse square of the inter-ring tunnelling in the weak-link limit. Number fluctuations then remain of order one instead of growing with the particle number, and within each number sector the system evolves into coherent superpositions of two many-body configurations, distinguished by which ring carries the circulation. The entanglement accessible under number-conserving operations therefore reaches up to approximately one bit. We propose a quench protocol to prepare the initial state and estimate transfer times within reach of current ring-lattice experiments, identifying coupled rings as a platform to study coherent quantum phase slips and to manipulate persistent-current states.
\end{abstract}

\section{Introduction}

Quantized circulation is a consequence of phase coherence in a closed geometry: single-valuedness of the order parameter 
constrains the
persistent-current states 
to live 
in topologically distinct sectors labeled by the winding number $\omega$.
In a classical-field description, changing the winding number requires the density to vanish locally, and therefore states with different winding numbers are separated by an energy barrier. Quantum mechanically, however, tunnelling between distinct winding-number sectors can occur, leading to a hybridization of the bare circulation states  \cite{hallwood2006macroscopic}. Understanding the microscopic mechanism underlying this tunnelling, and the many-body states that mediate transitions between different winding numbers, therefore provides a direct way to probe how the topological stability of persistent currents is overcome.
Such a change of winding number is known as a phase slip. Consequently, understanding the coherent transfer of circulation amounts to understanding the microscopic realization of these phase slips. In annular superfluids, this process is mediated by localized density defects that unwind the phase, such as vortices crossing the annulus in two-dimensional geometries \cite{yakimenko2015vortices} or solitonic excitations in effectively one-dimensional systems \cite{perez2020current}. Phase slips have been extensively studied in superconducting nanowires and circuits \cite{little1967decay}, where incoherent events contribute to the decay of persistent currents \cite{Arutyunov2008}, whereas coherent quantum phase slips hybridize states of different fluxoid number and generate coherent superpositions of distinct circulating-current states \cite{MooijHarmans2005,Astafiev2012}. From the perspective of superconducting circuits, coherent quantum phase slips constitute the electromagnetic dual of Cooper-pair tunnelling across a Josephson junction \cite{MooijNazarov2006,rastelli2013quantum}. 

Ultracold atomic rings provide a particularly controlled setting for this problem, as persistent currents can be prepared through synthetic gauge fields \cite{dalibard2011colloquium}, phase imprinting \cite{phase_imprinting}, or rotating weak links \cite{ramanathan2011superflow}, and their circulation can be directly detected in both bosonic and fermionic systems \cite{Ryu_2014,ring_review,polo2025persistent,DelPace2022}.
Beyond this experimental control, these systems allow the dynamics of quantized circulation to be isolated within a clean many-body Hamiltonian and characterized through observables that resolve particle-number fluctuations and coherence. They are therefore particularly well suited to distinguishing different microscopic mechanisms of coherent circulation transfer \cite{polkovnikov2005decay,derrico2017quantum}.

Considerable effort has been devoted to the preparation and stability of persistent currents in ultracold gases \cite{polo2025persistent}, including controlled generation of circulation \cite{Wright2013}, hysteretic dynamics in atomic rings \cite{Eckel2014}, and the decay of current states through thermal or quantum fluctuations \cite{Kumar2017,moulder2012quantized,polo2019oscillations}. In most of these settings the central process is the creation of a current state or its decay, schematically $0\rightarrow \omega$ or $\omega\rightarrow \omega-1$. A distinct operation is the coherent and reversible transfer of circulation between coupled subsystems,
where the current is 
transported between two quantum rings. 
The winding number can consequently be regarded not only as a signature of phase coherence, but also as a discrete degree of freedom that can be coherently manipulated.

Coherent circulation transfer has already been investigated at the mean-field level in a variety of coupled annular geometries. For laterally coupled rings, rotational fluxons and the conditions for their macroscopic tunnelling were identified early on \cite{brand2009rotational}, while introducing a tunable weak link between two density-coupled rings was later shown to produce controllable periodic transfer of the current, together with its suppression by dissipation and temperature \cite{Bland2022}. By contrast, without the weak link, quenches across the condensation transition were also found to generate persistent currents in two rings \cite{Bland2020}. Related transfer phenomena have also been studied in stacked or coaxial rings
\cite{oliinyk2019tunneling,nicolau2020orbital,Bazhan2022}, and between coherently coupled components of a spinor condensate confined in a single toroidal trap \cite{Gallemi2015}. In all these cases the dynamics is described within a classical-field picture, in which circulation is transported hydrodynamically through the evolution of the condensate density and phase. Even within this framework, however, identifying the microscopic mechanism has required considerable care: the observed oscillations have recently been attributed to low-energy collective sound modes rather than to the motion of an individual vortex core \cite{chaika2026acoustic}. By construction, however, mean-field descriptions cannot resolve the underlying many-body dynamics.

Beyond the mean-field approximation, previous studies have investigated the effective low-energy dynamics, vortex-like excitations, vortex tunnelling, and static properties of coupled rings~\cite{aghamalyan2013effective,richaud2017quantum,escriva2019tunneling,escriva2021static}. 
In addition, the symmetries of laterally coupled rings have been shown to make the tunnelling amplitudes between different winding states complex-valued~\cite{polo2016geometrically}. A Bose–Hubbard study of two coupled rings further demonstrated that a persistent current can be coherently transferred while the average particle populations remain approximately balanced, identifying this transfer as a coherent quantum phase slip (QPS) process~\cite{PerezObiol2022}. 
Although quantum phase slips were previously established as a mechanism for circulation exchange, this was shown for few particles systems at vanishing or very strong interactions, and the associated entanglement was not quantified. Here, we extend the analysis to larger particle numbers, allowing us to explore two complementary limiting descriptions: a mean-field regime at weak interactions and a phase dominated regime at stronger interactions, both more closely connected to experimentally relevant conditions. Characterizing the entanglement in these regimes 
is particularly important because circulation transfer can also occur within mean-field theory, without involving quantum phase slips.

A given transfer of circulation may originate either from the coherent exchange of particles occupying well-defined momentum modes or from the coherent hybridization of collective many-body current states localized in opposite rings. Existing studies have largely focused either on the mean-field limit~\cite{brand2009rotational,Bland2022,oliinyk2019tunneling,nicolau2020orbital,Bazhan2022,Gallemi2015,chaika2026acoustic} or on specific many-body regimes~\cite{aghamalyan2013effective,richaud2017quantum,escriva2019tunneling,escriva2021static,PerezObiol2022}, leaving open whether coherent circulation transfer is governed by a single microscopic mechanism or by qualitatively distinct physical processes. Distinguishing these possibilities requires going beyond the average current and resolving particle-number fluctuations, condensation properties, and the structure of the entanglement generated during the transfer.

To distinguish between these microscopic scenarios, we investigate coherent persistent-current transfer between two weakly coupled lattice rings within a unified Bose–Hubbard description at large filling, where both the Gross–Pitaevskii and Quantum Phase Models (QPM) emerge as controlled effective theories in their respective regimes of validity. We show that remarkably similar oscillations of the average current can arise from qualitatively different many-body dynamics. We find that these mechanisms are distinguished by different scalings of the transfer time with the inter-ring tunnelling, as well as by their particle-number fluctuations, condensate fragmentation, and entanglement structure. 
Importantly, these mechanisms belong to distinct interaction regimes and are not continuously connected: the mode exchange dynamics is ultimately suppressed by self-trapping, whereas the collective phase-dominated mechanism emerges in the quantum-phase regime.

\section{Physical picture}

\begin{figure}[h]
    \centering
    \includegraphics[width=1\linewidth]{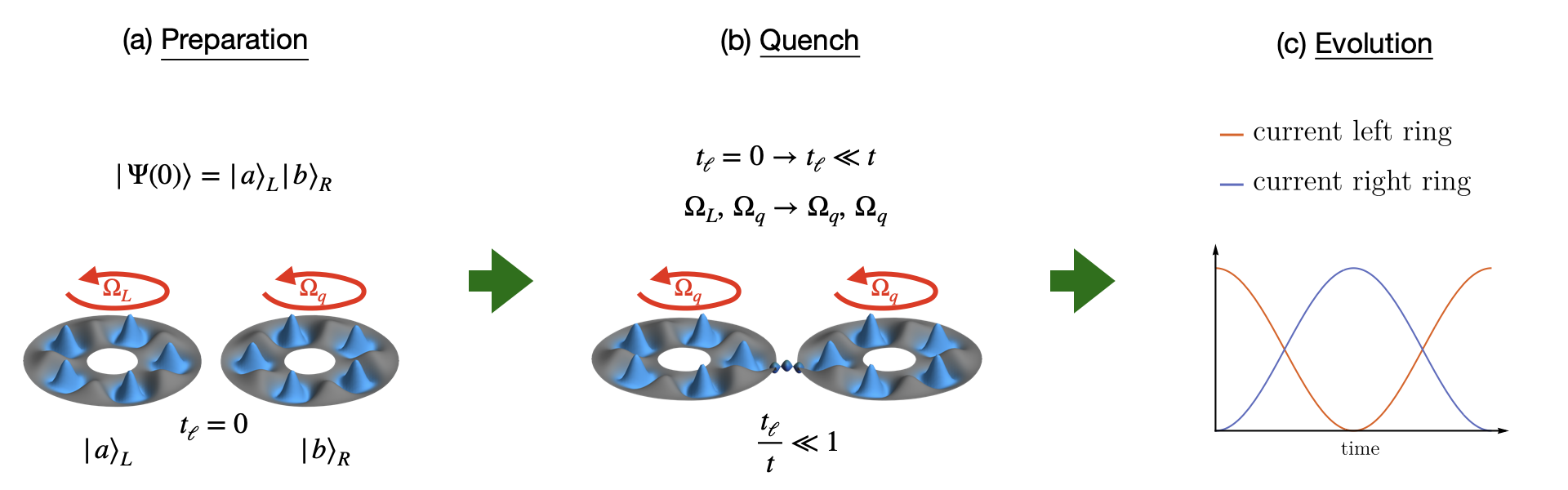}
    \caption{Schematic illustration of the physical system and quench protocol. (a) Preparation: Two identical, initially uncoupled rings ($t_\ell=0$) are loaded with the same fixed number of particles in each ring. Different synthetic fluxes, $\Omega_L$ and $\Omega_q$, are applied to the left and right rings, respectively. The system is allowed to relax to its ground state, thereby preparing different persistent currents in the two rings. (b) Quench: The rings are suddenly coupled, for example by lowering the inter-ring barrier or bringing the rings into close proximity, resulting in a finite but weak interring tunnelling, $t_\ell>0$. At the same time, the left-ring flux is quenched from $\Omega_L$ to $\Omega_q$, so that both rings experience the same flux after the quench. (c) Evolution: The system evolves freely under the post-quench Hamiltonian, giving rise to coherent oscillations of the average current between the two rings.}
    \label{fig:protocol}
\end{figure}

We consider two side-by-side rings, with 
one unit of circulation (winding number) initially prepared in one ring while the other is in the zero-winding state.
We then couple the rings and, under certain conditions, the circulation starts to coherently transfer back and forth between them while their average populations remain essentially unchanged. Remarkably, such current oscillations can arise from two qualitatively different microscopic mechanisms.

In the weakly interacting regime, each occupied momentum mode forms an effective two-level system, with the two levels corresponding to the same momentum state in the two rings. Each particle undergoes Rabi-like oscillations between the rings while remaining in that momentum mode, so that the many-body dynamics is assembled from independent single-particle tunnelling processes. This mode-exchange dynamics is closely related to the familiar two-mode Josephson problem, but with the two effective modes corresponding to current states localized in the two rings: increasing the interaction strength drives the system from Rabi-like oscillations toward nonlinear Josephson dynamics and, eventually, current self-trapping, which suppresses complete transfer. 
At stronger interactions, where the mean-field description ceases to apply, a qualitatively distinct mechanism emerges. In this phase-dominated regime, particle-number fluctuations are energetically suppressed and the dynamics is largely confined to a small many-body subspace formed by collective current configurations. The current is then transferred through the coherent hybridization of these collective states rather than through the tunnelling of particles within individual momentum modes, giving rise to coherent superpositions of distinct current states with a NOON-like structure in the current degree of freedom.

\section{Models and observables}

\subsection{Bose--Hubbard}

We describe the system of two weakly coupled lattice rings, by the Hamiltonian
\begin{eqnarray}
&&H_{\mathrm{BH}}=H_L+H_R+H_I,\nonumber\\
&&H_\alpha=
\sum_{i=0}^{N_s-1}
\left[
\frac{U}{2}
(\hat n_{\alpha,i}^2-\hat n_{\alpha,i})
-t\left(e^{-i\frac{2\pi\Omega_\alpha}{N_s}}\hat a_{\alpha,i}^\dagger\hat a_{\alpha,i+1}+h.c.\right)
\right],
\nonumber
\\
&&H_I=
-t_\ell
\left(
\hat a_{L,0}^\dagger\hat a_{R,0}
+\hat a_{R,0}^\dagger\hat a_{L,0}
\right),
\label{eq:HBH}
\end{eqnarray}
where
$\hat a_{\alpha,i}$ annihilates a boson on site $i$ of ring $\alpha$,
$\hat n_{\alpha,i}=\hat a_{\alpha,i}^\dagger\hat a_{\alpha,i}$,
and periodic boundary conditions
$\hat a_{\alpha,N_s}=\hat a_{\alpha,0}$
are assumed. 

The persistent current is obtained from the derivative of the Hamiltonian with respect to the synthetic flux,
$J_\alpha=\langle\hat J_\alpha\rangle=-N_s/(2\pi\hbar)\langle\partial H_\alpha/\partial\Omega_\alpha\rangle$, which yields
\[
J_\alpha=\frac{2t}{\hbar}\sum_{i=0}^{N_s-1}\mathrm{Im}\left[e^{-i\frac{2\pi\Omega_\alpha}{N_s}}
\langle\hat a_{\alpha,i}^\dagger\hat a_{\alpha,i+1}\rangle\right].
\]

To characterize particle-number fluctuations, we define the total number operator
$\hat N_\alpha=\sum_i \hat n_{\alpha,i}$,
together with the average ring and site variances,
\begin{align}
\sigma_r^2
&=
\frac{1}{2}
\sum_{\alpha=L,R}
\left(
\langle\hat N_\alpha^2\rangle
-
\langle\hat N_\alpha\rangle^2
\right),
\\
\sigma_s^2
&=
\frac{1}{2N_s}
\sum_{\alpha=L,R}\sum_i
\left(
\langle\hat n_{\alpha,i}^2\rangle
-
\langle\hat n_{\alpha,i}\rangle^2
\right).
\end{align}

The condensate fraction is obtained from the largest eigenvalue
$\lambda_{\max}$
of the single-particle density matrix
$
\rho^{(1)}_{(\alpha,i),(\beta,j)}
=
\langle
\hat a_{\alpha,i}^\dagger
\hat a_{\beta,j}
\rangle
$,
through
$f_c=\lambda_{\max}/N_p$.

We quantify entanglement between the left and right rings by performing a Schmidt decomposition of the combined many-body state, treating each ring as a subsystem;
$
|\Psi\rangle
=
\sum_k
\sqrt{\lambda_k}\,
|\phi_k^{(L)}\rangle
|\phi_k^{(R)}\rangle
$,
where Schmidt coefficients $\lambda_k$ define the von Neumann entropy
$
S=
-\sum_k
\lambda_k
\log_2\lambda_k$. 
Because the total particle number is fixed, the reduced density matrix is
block diagonal in the particle number \(n\) of the left ring. Its entropy
can therefore be decomposed as
\[
S
=
-\sum_n P_n\log_2P_n
+
\sum_nP_nS_n
\equiv
S_N+S_{\rm acc},
\]
where \(P_n\) is the probability of finding \(n\) particles in the left
ring and \(S_n\) is the entropy of the normalized state projected onto
that number sector. The first contribution, \(S_N\), is the Shannon
entropy associated with ring-number fluctuations. The second,
\(S_{\rm acc}\), is the entanglement accessible under local operations
that conserve particle number.
Within the most probable particle-number sector, we additionally define $
W_2=\lambda_1^{(\mathrm{dom})}
+\lambda_2^{(\mathrm{dom})}$,
where the \(\lambda_i^{(\mathrm{dom})}\) are the normalized Schmidt
probabilities conditioned on that sector. This quantity diagnoses whether the entanglement within the dominant
number sector is effectively two dimensional.

\subsection{Mean field limits: discrete and continuous Gross--Pitaevskii equations}

In the mean-field limit, the Bose--Hubbard Hamiltonian Eq.~\ref{eq:HBH} reduces to two coupled discrete Gross--Pitaevskii equations by replacing the bosonic operators with their expectation values,
$\hat a_{\alpha,j}\rightarrow\langle\hat a_{\alpha,j}\rangle\equiv\psi_{\alpha,j}$,
and neglecting quantum fluctuations. The resulting equations for the coupled lattice rings are
\begin{align}
i\hbar\,\dot\psi_{L,j}
&=
- t \left(
e^{-i\frac{2\pi\Omega_L}{N_s}}\psi_{L,j+1}
+ e^{i\frac{2\pi\Omega_L}{N_s}}\psi_{L,j-1}
\right)
+ U\,|\psi_{L,j}|^2\psi_{L,j}
- t_\ell\,\psi_{R,0}\,\delta_{j0},
\\
i\hbar\,\dot\psi_{R,j}
&=
- t \left(e^{-i\frac{2\pi\Omega_R}{N_s}}\psi_{R,j+1}
+ e^{i\frac{2\pi\Omega_R}{N_s}}\psi_{R,j-1}\right)
+ U\,|\psi_{R,j}|^2\psi_{R,j}
- t_\ell\,\psi_{L,0}\,\delta_{j0}.
\end{align}
We normalize the discrete fields as $\sum_{j=0}^{N_s-1} |\psi_{L,j}(t)|^2=\sum_{j=0}^{N_s-1} |\psi_{R,j}(t)|^2=N_p/2$. 
The mean field current on ring $\alpha=L,R$ is
\[
J_\alpha=\frac{2t}{\hbar}\,\sum_{j=0}^{N_s-1} 
\mathrm{Im}
\left[
e^{-i\frac{2\pi\Omega_\alpha}{N_s}}
\psi_{\alpha,j}^*\psi_{\alpha,j+1}
\right].
\]

In the continuum limit, the mean-field dynamics is described by two coupled one-dimensional Gross--Pitaevskii equations on rings of circumference $L$,
\begin{align}
i\hbar\,\partial_t \psi_L(x,t)
&=
\frac{\hbar^2}{2m}\left(-i\partial_x-\frac{2\pi\,\Omega_L}{L}\right)^2\psi_L
+g|\psi_L|^2\psi_L
-t_\ell\,f(x)\,\psi_R(x,t),
\\
i\hbar\,\partial_t \psi_R(x,t)
&=
\frac{\hbar^2}{2m}\left(-i\partial_x-\frac{2\pi\,\Omega_R}{L}\right)^2\psi_R
+g|\psi_R|^2\psi_R
-t_\ell\,f(x)\,\psi_L(x,t).
\end{align}
Here $f(x)$ is a normalized Gaussian centered at $x=0$,
$\frac{1}{L}\int_0^{L} dx\,f(x)=1$,
which models the localized inter-ring tunnelling region.
Initially, each condensate is normalized according to
$\int_0^{L} dx\,|\psi_{L,R}(x,0)|^2=1$.
For a population of $N_p/2$ atoms per ring, the nonlinear interaction strength is
$g=(N_p/2)g_{1D}$,
where
$g_{1D}=2\hbar^2a_s/(ma_\perp^2)$, with $a_s$ the $s$-wave scattering length and $a_\perp$ the transverse harmonic oscillator length.
The persistent current in each ring is given by
\[
J=\frac{\hbar}{m}\int_0^{L}dx\left(\mathrm{Im}[\psi^*\partial_x\psi]-\frac{2\pi\Omega}{L}|\psi|^2\right).
\]

In the weakly interacting regime, the coupled GPEs can be further reduced to an effective two-mode model, derived in Appendix~\ref{app:twomode}.

\subsection{Quantum phase model}

The QPM is obtained from the Bose--Hubbard model in the large-filling, phase-dominated regime by writing $\hat n_{\alpha,i}=\rho+\delta\hat n_{\alpha,i}$, where $\rho$ is a site-independent average occupation, taken to be constant in space and time at the BH level, and $\delta\hat n_{\alpha,i}$ describes number fluctuations around it. For two rings containing $N_p/2$ atoms each, one has $\rho=N_p/(2N_s)$. We label the maximum fluctuation considered as $N_{\rm max}$. We show how the QPM is derived from the BHM in Appendix~\ref{app:qpm}. Neglecting amplitude fluctuations, the Bose--Hubbard hopping terms with tunnelling amplitudes $t$ and $t_\ell$ reduce to phase-only Josephson couplings, while the interaction term retains the same coefficient,
\[
t_{\mathrm{QPM}}=\rho\,t,
\qquad
t_\ell^{\mathrm{QPM}}=\rho\,t_\ell,
\qquad
U_{\mathrm{QPM}}=U.
\]
\begin{align}
H_{\mathrm{QPM}}&=H_L^{\mathrm{QPM}}+H_R^{\mathrm{QPM}}+H_I^{\mathrm{QPM}},
\\
H_\alpha^{\mathrm{QPM}}
&=
\sum_{i=0}^{N_s-1}
\left[
\frac{U_{\mathrm{QPM}}}{2}\,\delta \hat n_{\alpha,i}^{\,2}
-
2\,t_{\mathrm{QPM}}
\cos\!\left(
\hat{\phi}_{\alpha,i+1}-\hat{\phi}_{\alpha,i}
-\frac{2\pi\Omega_\alpha}{N_s}
\right)
\right],
\\
H_I^{\mathrm{QPM}}
&=
-\,2\,t_{\ell}^{\mathrm{QPM}}
\cos\!\left(
\hat{\phi}_{R,0}-\hat{\phi}_{L,0}
\right),
\end{align}
with $[\delta\hat n_{\alpha,i},\hat\phi_{\beta,j}]=i\delta_{\alpha\beta}\delta_{ij}$.

The total current on each ring is defined consistently with the Bose–Hubbard model as
$
\hat J_\alpha^{\mathrm{QPM}}=
-N_s/(2\pi\,\hbar)
\frac{\partial H_\alpha^{\mathrm{QPM}}}{\partial\Omega_\alpha},
$
which gives
\[
\hat J_\alpha^{\mathrm{QPM}}
=
\frac{2\,t_{\mathrm{QPM}}}{\hbar}
\sum_{i=0}^{N_s-1}
\sin\!\left(
\hat\phi_{\alpha,i+1}-\hat\phi_{\alpha,i}
-\frac{2\pi\Omega_\alpha}{N_s}
\right).
\]

\subsection{Quench protocol}
\label{sec:protocol}

The protocol considered here provides a convenient route to prepare a persistent circulation localized in one ring and subsequently trigger the coherent current transfer we are investigating. It is formulated in terms of ground-state preparation followed by a quench, but this particular sequence is not essential: other preparation schemes could in principle be used to realize the same initial current configuration, which may be more suitable/convenient experimentally, e.g. phase imprinting, dynamical protocols or other quenches~\cite{polo2025persistent}.

The main idea of the protocol is to prepare a state with circulation localized in one ring that, after the quench, overlaps with a small set of hybridized eigenstates of the coupled system. After the quench, both rings experience the same nonzero flux, $\Omega_q$, which breaks the clockwise/counterclockwise symmetry while preserving exchange symmetry between the rings. In the uncoupled limit, exchange-related states are degenerate; weak inter-ring tunnelling lifts this degeneracy by hybridizing them into delocalized eigenstates. If the prepared state overlaps predominantly with the resulting hybridized eigenstates, their interference produces coherent oscillations of the persistent current between the two rings. This motivates both the choice of a common post-quench flux $\Omega_q$ and the preparation of a state  with finite winding initially localized in only one ring.

Specifically, we consider two identical rings initially uncoupled ($t_\ell=0$) and with $N_p/2$ bosons each, and prepare their ground state at different synthetic fluxes $\Omega_L$ and $\Omega_R$. At time $\tau=0$, the system is quenched by switching on a weak inter-ring tunnelling, $t_\ell\ll t$, while simultaneously setting the fluxes in both rings to $\Omega_q\neq0$.

The specific implementation depends on the interaction regime. In the mode-exchange regime ($U\rho/t\ll1$), the single-particle eigenstates are independent of the applied flux, which only shifts their energies. Varying the flux therefore changes which momentum eigenstate minimizes the energy, without modifying the eigenstates themselves. As a result, both winding-one and zero-winding states can be straightforwardly prepared as ground states of isolated rings by choosing suitable initial fluxes $\Omega_L$ and $\Omega_R$.

In the phase-dominated regime ($U/(\rho t)=\mathcal O(1)$), interactions substantially modify the many-body eigenstates of the uncoupled system as the flux is varied, making the preparation of the desired winding-localized state less straightforward. 
The preparation flux in the first ring, $\Omega_L$, is therefore determined numerically by maximizing the overlap between the single-ring many-body ground state at $\Omega_L$ and the target current-carrying eigenstate at the post-quench flux $\Omega_q=\Omega_R$. The corresponding overlap maps for different interaction strengths and lattice sizes are shown in Appendix~\ref{app:optimization}.

Since these maps are computed for uncoupled rings ($t_\ell=0$), they provide an upper bound for the actual preparation overlap after the quench. Finite inter-ring tunnelling hybridizes the product states and mixes additional many-body states, so the actual preparation overlap is generally slightly smaller, with the difference increasing as the system moves away from the perturbative regime.

For lattice rings of $N_s=3$ sites, the optimized overlaps reach approximately $0.95$ for $U/(\rho t)=0.5$ and $0.8$ for $U/(\rho t)=1.5$, which are large enough to produce long-lived coherent current oscillations. Increasing the number of sites to $N_s=5$ further improves these values to $0.99$ and $0.93$, respectively.

\section{Persistent current transfer}

\subsection{Representative current-transfer dynamics}
\label{sec:ctdynamics}

\begin{figure}[th]
    \centering
\includegraphics[width=1\linewidth]{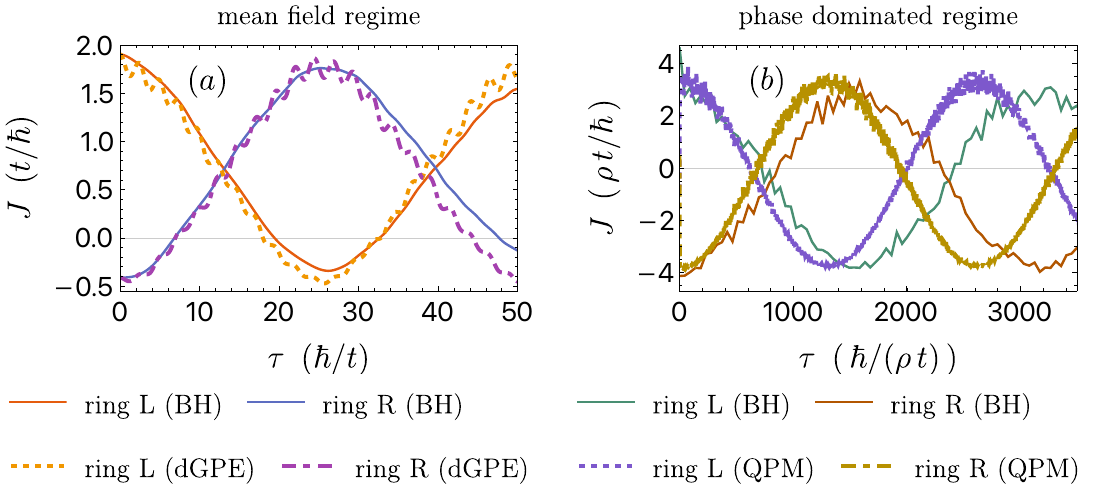}
    \caption{
    Current transfer between left and right rings in the mean field regime (plot (a) on the left), and phase dominated regime (plot (b) on the right). (a) Simulations using rings of $N_s=3$ sites, with $N_p=42$ bosons, $U=0.02$, $t_\ell=0.2$, $\Omega_L=1$, $\Omega_R=0.1$, $\Omega_q=0.1$. (b) BH simulations using $N_s=3$, $N_p=42$, $U=3.5$, $t_\ell=0.1$, $\Omega_L=0.55$, $\Omega_R=\Omega_q=0.46$. For the QPM, we have the same values, corresponding to $U_{\rm QPM}=0.5$, and used a fluctuation cutoff $N_{\rm max}=3$. }
    \label{fig:universal_current}
\end{figure}

We now turn to the post-quench dynamics of two weakly coupled rings following the protocol introduced in Sec.~\ref{sec:protocol}. Figure~\ref{fig:universal_current} summarizes the current evolution in the two interaction regimes studied throughout this work: the mode-exchange regime, $U\rho/t\ll1$, and the phase-dominated regime, $U/(\rho t)=\mathcal O(1)$. In both cases the persistent current is coherently transferred between the two rings while the average population of each ring remains essentially constant, $\langle\hat N_L\rangle=\langle\hat N_R\rangle\approx N_p/2$. The microscopic origin of this transfer, however, is fundamentally different in the two regimes and will be analyzed in the following subsections.

The simulations correspond to two rings of $N_s=3$ sites containing a total of $N_p=42$ bosons, corresponding to a filling $\rho=7$. Throughout the evolution we compute the ring currents together with the ring and site occupations and their corresponding fluctuations.

Panel (a) shows the weakly interacting mode-exchange regime. The Bose--Hubbard and discrete Gross--Pitaevskii simulations are performed with $U/t=0.02$, corresponding to $U\rho/t=0.14\ll1$. The initial state is the ground state of two uncoupled rings ($t_\ell=0$) with fluxes $\Omega_L=1$ and $\Omega_R=0.1$, corresponding approximately to a plane wave carrying a persistent current in the left ring and a zero-winding condensate in the right ring, which carries only a small current $J_0=-(2t/\hbar)\sin(2\pi\Omega_q/N_s)\approx-0.42\,t/\hbar$ per particle due to the flux. The quench consists of switching on the coupling to $t_\ell/t=0.2$ while setting both fluxes to $\Omega_q=0.1$. Both models predict very similar dynamics throughout the evolution, exhibiting a complete transfer of the persistent current after $\tau=25\,(\hbar/t)$. The currents oscillate approximately sinusoidally between $-0.5\,(t/\hbar)$ and $2\,(t/\hbar)$, per particle, with period $T_{\rm MF}\simeq 50\,(\hbar/t)$. The expectation value of the particle number in each ring is the same throughout the evolution.

Panel (b) shows the phase-dominated regime. Here we compare Bose--Hubbard and QPM simulations for $U/t=3.5$, $N_p=42$, corresponding to $U/(\rho t)=0.5$, using a QPM cutoff $N_{\rm max}=3$, for which all observables are well converged. The initial state is prepared as the ground state of two uncoupled rings with $\Omega_L=0.55$ and $\Omega_R=0.46$, while the post-quench Hamiltonian is obtained by setting $t_\ell/t=0.1$ and equal fluxes $\Omega_q=0.46$. As discussed in Sec.~\ref{sec:protocol}, these fluxes maximize the overlap between the initial ground state and the relevant post-quench current doublet. Both models again predict coherent oscillations of the persistent current between the rings. The currents evolve approximately sinusoidally between $-4\,(\rho t)/\hbar$ and $4\,(\rho t)/\hbar$ with a period of approximately $T_{\rm PD}\simeq3000\,\hbar/(\rho t)$. The agreement is slightly less quantitative than in the mode-exchange regime, with the Bose--Hubbard model predicting a somewhat slower oscillation together with small additional fluctuations, but both descriptions reproduce the same oscillation amplitude, timescale, and overall dynamics. In this regime, particle number in each ring is also balanced during the full evolution.

The excellent agreement between the Bose--Hubbard model and the corresponding effective descriptions in both interaction regimes demonstrates that the discrete Gross--Pitaevskii equations accurately describe the weakly interacting limit, while the quantum phase model captures the phase-dominated regime. The microscopic mechanisms responsible for the coherent current transfer in each case are analyzed in the following subsections.

\subsection{Weak interaction}
The dynamics in the mode-exchange regime can be understood analytically by considering the non-interacting limit, $U=0$, derived in Appendix~\ref{app:bh}. We first solve the single-particle problem, corresponding to one particle occupying a momentum state in each ring, and then generalize the solution to $N_p/2$ bosons per ring by exploiting the fact that, in this limit, the bosons evolve independently. After Fourier transforming the Bose-Hubbard Hamiltonian and treating the interring tunnelling perturbatively, the dynamics factorizes into independent momentum sectors, each described by a two-level system connecting the corresponding momentum states in the two rings. The initial state is therefore a product state in the ring degree of freedom, with every occupied momentum mode localized in one ring. Therefore, the left-right entanglement vanishes and the condensate fraction of each ring is initially one.

Following the quench, particles occupying each momentum mode undergo independent Rabi oscillations between the two rings while remaining in the same momentum mode. At $\tau=T/4$, every mode occupies an equal coherent superposition of the left and right rings. In the corresponding many-body Bose--Hubbard description, this is also the instant at which the ring-number fluctuations and left-right entanglement reach their maximum values, while the condensate fraction of each ring is reduced to $f_c^{\rm ring}=1/2$, since each ring is equally populated by two orthogonal momentum modes. For $N_p$ bosons, the many-body dynamics follows directly from the single-particle solution in this noninteracting limit, as the particles evolve independently. In particular, the oscillation period is set by an energy gap that scales linearly with the inter-ring coupling $t_\ell$,
$T=\frac{\pi N_s}{t_\ell}$, which remains independent of particle number, while the average occupation of each ring is conserved, $\langle N_{L,R}\rangle=N_p/2$. The ring-number variance grows extensively with the system size,
$\overline{\sigma}_{N_{L,R}}^2=\frac{N_p}{8}$,
or equivalently $\overline{\sigma}_{N_{L,R}}^2/(N_p/2)=1/4$, demonstrating that the fluctuations scale linearly with the total number of particles. For the site-number variance, the non-interacting limit predicts
$\overline{\sigma}_s^2=\rho+\frac{\rho^2}{4}-\frac{3\rho}{4N_s}$.
The global condensate fraction remains constant throughout the evolution at $f_c^{\rm tot}=1/2$, whereas the condensate fraction of each individual ring oscillates between $1$ and $1/2$. 
The total and accessible left-right entanglement both grow logarithmically with particle number. In particular, even after conditioning on the ring population, the mode-exchange state retains a higher-dimensional entanglement associated with the possible distributions of the particles among the two occupied momentum modes. The corresponding analytical expressions are given in
Appendix~\ref{app:bh}.

The numerical Bose-Hubbard simulations shown in Fig.~\ref{fig:universal_current}(a), corresponding to $U\rho/t=0.14$, are remarkably close to these analytical predictions despite the presence of weak interactions. Averaging over one complete oscillation period yields $\overline{\sigma_r^2}/(N_p/2)=0.26$ and $\overline{\sigma_s^2}/\rho=2.8$, in good agreement with the non-interacting values, corresponding to 0.25 and 2.5. The maximum von Neumann entanglement entropy reaches $S_{vN}^{\rm max}=5.8$ at $\tau=T/4$, while the condensate fraction of each ring oscillates between $1$ and $1/2$ and the total condensate fraction remains constant at $1/2$ throughout the evolution. Finally, simulations performed for $N_p=30,36,42,$ and $48$ confirm the predicted scaling of the particle-number fluctuations with system size, as summarized in Table~\ref{tab:appendixResults} of Appendix~\ref{app:sims}.

\subsection{Strong interaction}

In the phase-dominated regime, the transfer mechanism is fundamentally different from the mode-exchange picture discussed above. Because transferring particles between the rings carries a large interaction-energy cost, the dynamics remains confined to the particle-number sectors around the balanced configuration $N_L=N_R=N_p/2$. Current transfer is therefore not accompanied by a macroscopic exchange of particles, but it results from the coherent hybridization of many-body persistent-current states.

This behavior is clearly reflected in the particle-number fluctuations. For the simulation shown in Fig.~\ref{fig:universal_current}(b), we obtain time-averaged values of $\overline{\sigma_r^2}/(N_p/2)=0.02$ and $\overline{\sigma_s^2}/\rho=0.11$, over an order of magnitude smaller than in the mean-field regime. Consequently, the wavefunction remains almost entirely confined to only a few particle-number sectors centered around $N_L=N_R=N_p/2$.

The entanglement structure provides an additional signature of the collective current-transfer mechanism. We denote by $|N_p/2,0\rangle$ and $|N_p/2,1\rangle$ two orthogonal many-body eigenstates of an isolated ring at the post-quench flux, both containing $N_p/2$ particles but carrying different persistent currents. Here, the second label distinguishes the two current-carrying many-body states and does not refer to a particle occupation number. Both states may include interaction-induced correlations within the ring. In the ideal weak-link limit, the dynamics is restricted to the configurations $|N_p/2,1\rangle_L|N_p/2,0\rangle_R$ and $|N_p/2,0\rangle_L|N_p/2,1\rangle_R$. Starting from the former, their resonant coupling gives
\[
|\Psi(\tau)\rangle=
\cos\theta\,|N_p/2,1\rangle_L|N_p/2,0\rangle_R
-i\sin\theta\,|N_p/2,0\rangle_L|N_p/2,1\rangle_R,
\qquad
\theta=\frac{\Delta E\,\tau}{2\hbar},
\]
up to an overall phase, where $\Delta E$ is the splitting of the hybridized doublet.

Since the local states are orthogonal, the two nonzero Schmidt probabilities are \(\cos^2\theta\) and \(\sin^2\theta\). When ordered from largest to smallest, they read
\[
\lambda_1=\max\{\cos^2\theta,\sin^2\theta\},
\qquad
\lambda_2=\min\{\cos^2\theta,\sin^2\theta\},
\]
and therefore \(W_2=\lambda_1+\lambda_2=1\) throughout the ideal evolution. At the first current crossing, \(\theta=\pi/4\), both probabilities equal \(1/2\), and the state becomes a maximally entangled Bell state of the two collective current degrees of freedom. Because both configurations have the same particle number in each ring, the number entropy vanishes and the full one bit of left–right entanglement is accessible under local particle-number-conserving operations.

This state can be viewed as a current-state analogue of a NOON superposition: the two branches exchange the collective current configurations while preserving the local atom numbers. The analogy concerns the collective current degree of freedom, rather than a superposition with all atoms occupying either ring.

For the finite couplings and preparation fidelities used in the simulations, neighbouring particle-number sectors and additional many-body eigenstates modify this ideal evolution. Within the dominant number sector, the two leading Schmidt probabilities approximately follow the ordered two-state pattern, with additional temporal modulations and without necessarily becoming exactly equal at the current crossing. Nevertheless, they account for almost the entire conditional Schmidt weight, \(W_2\simeq1\), while the accessible entanglement remains close to one bit near the crossing. The overlaps with the two product configurations likewise support their dominant role, although the evolving state differs appreciably from the ideal two-state superposition. Together, these results identify an approximately two-dimensional collective entanglement structure, rather than an exact NOON-like state.

The post-quench spectral splitting in the phase dominated regime scales as
$\Delta E_{\rm PD}\propto t_\ell^2$,
leading to the characteristic current-transfer period
$T\propto t_\ell^{-2}$.
Unlike the linear scaling found in the mean-field regime, this quadratic dependence is the hallmark of a collective second-order tunnelling process.

The dependence on particle number follows directly from the QPM description. Since $t_{\rm QPM}=\rho\,t$, the dimensionless dynamics depends only on the ratios $U/(\rho t)$ and $t_\ell/t$. Keeping these parameters fixed therefore gives
$T=\frac{\hbar}{\rho t}
\widetilde T
\!\left(
\frac{U}{\rho t},
\frac{t_\ell}{t}
\right)
\propto
\rho^{-1}
\propto
N_p^{-1}$,
while the ring- and site-number fluctuations remain approximately independent of $N_p$. The numerical results obtained for $N_p=30,36,42,$ and $48$, summarized in Table~\ref{tab:appendixResults}, fully support these predicted scalings.

\subsection{Comparison of the two regimes }
The results presented above demonstrate that coherent persistent-current transfer can emerge from two qualitatively distinct microscopic mechanisms. Although the current dynamics is similar at the level of expectation values, the underlying many-body physics is fundamentally different. In the mode-exchange regime, the many-body evolution is built from several independent single-mode tunnelling processes and is not confined to an isolated many-body doublet. In the phase-dominated regime, by contrast, the spectral weight is concentrated predominantly in a quasi-degenerate pair of collective current states. Table~\ref{tab:regimesDiff} summarizes the main signatures that distinguish the two regimes.

\begin{table}[th]
\centering
{\renewcommand{\arraystretch}{1.6}
\begin{tabular}
{l|c|c}
 & Mode-exchange (mean field) & Phase-dominated 
 \\
\hline
Interaction regime
& $\frac{U\rho}{t}\ll1,\quad
\frac{U\rho}{t_\ell}<\frac{4}{N_s}$
& $\frac{U}{\rho\, t}=\mathcal O(1)$
\\[2pt]\hline
Current-transfer period
& $T\propto \frac{1}{t_\ell}$
& $T\propto \frac{1}{t_\ell^2\,N_p}$
\\[4pt]  \hline
Ring-number fluctuations
& $\overline{\sigma}_r^2\simeq \frac{N_p}{8}$
& $\overline{\sigma}_r^2\approx 0.5$
\\[2pt] \hline
Site-number fluctuations
& $\overline{\sigma}_s^2\simeq\rho+\frac{\rho^2}{4}-\frac{3\rho}{4N_s}$
& $\overline{\sigma}_s^2\lesssim 0.9$
\\[3pt] \hline
Two-Schmidt weight
& $W_2\lesssim 0.5$
& $W_2\approx 0.999$
\end{tabular}}
\caption{
Comparison of the characteristic properties of the mean-field and phase-dominated transport regimes. Overbars denote averages over one complete current-transfer period. The analytical estimates shown are supported by numerical Bose-Hubbard simulations and are representative of all particle numbers considered. Detailed numerical values for $N_p=30$, 36, 42, 48 are provided in Appendix~\ref{app:sims}.
 }
\label{tab:regimesDiff}
\end{table}

The most notable difference is the role of particle-number fluctuations. In the mode-exchange regime, current transfer is mediated by real particle tunnelling and the associated fluctuations grow extensively with the number of particles. In contrast, the phase-dominated regime suppresses real particle transfer, confining the dynamics to a few neighbouring number sectors while the current is transferred through collective phase dynamics. These different microscopic mechanisms also lead to distinct transfer-time scalings, $T\propto t_\ell^{-1}$ and $T\propto t_\ell^{-2}$, respectively.
The two mechanisms also generate qualitatively different entanglement structures. Mode exchange distributes the accessible left–right entanglement over multiple Schmidt components associated with different occupations of the momentum modes. In the phase-dominated regime, the dominant number sector instead exhibits an approximately two-dimensional Schmidt structure, associated predominantly with the two exchanged collective current configurations. Its ideal limit is a Bell state of the ring-current degrees of freedom, or equivalently a current-state analogue of a NOON superposition, with fixed atom number in each ring. The distinguishing feature is thus the collective encoding and concentration of entanglement in two current configurations, rather than a larger entanglement entropy.

\subsection{Validity of the mode-exchange regime }
\label{sec:mode-exchange-validity}

Although the microscopic mechanisms are now well understood, the results presented so far correspond to representative parameters chosen within the perturbative regime. An important question is therefore how far these mechanisms remain valid when the system parameters are varied. We first examine the domain of validity of the mode-exchange mechanism. Within the mean-field regime, increasing the interaction strength (or decreasing the inter-ring tunnelling) eventually drives a transition from complete mode exchange to mode self-trapping. Importantly, this self-trapping transition occurs entirely within the mean-field regime and is unrelated to the phase-dominated mechanism discussed above; the two transport regimes are separated by a parameter region in which neither mechanism produces coherent full current transfer.

We simulate two weakly coupled continuous rings using the coupled Gross--Pitaevskii equations and the same preparation and quench protocol introduced in Sec.~\ref{sec:protocol}. Continuous rings are used because they additionally reveal the spatial density evolution during the transfer. The transition itself, however, originates from the momentum-mode dynamics and therefore has a direct counterpart in the discrete GPE.

\begin{figure}[!htb]
    \centering
    \includegraphics[width=1\linewidth]{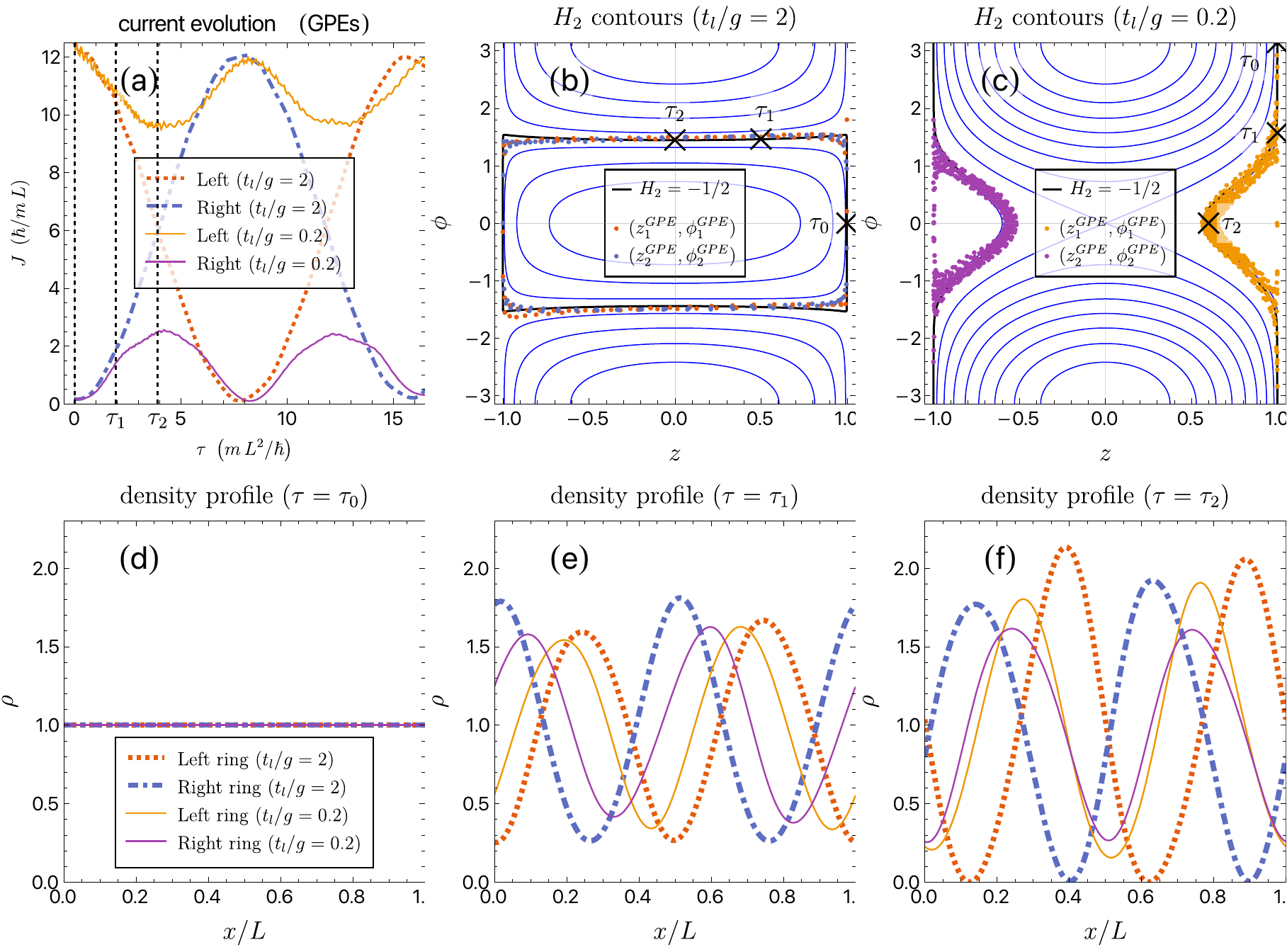}
    \caption{Two-mode dynamics in the mean-field (MF) regime. Comparison between the full-transfer regime ($t_\ell/g=2$, dashed lines) and the self-trapping regime ($t_\ell/g=0.2$, solid lines). The two ring states are prepared with winding numbers $q = 2$ (left) and $q = 0$ (right) and $2\pi\Omega/L=-0.15$. (a) Post-quench evolution of the average current in each ring over one oscillation period for both regimes, obtained from coupled Gross–Pitaevskii equation (GPE) simulations. (b,c) Constant-energy contours in the $(z,\phi)$ phase space for the full-transfer and self-trapping regimes, respectively. The thick black curve denotes the energy contour corresponding to the initial state ($|z|=1$). The filled circles show the values of $(z,\phi)$ extracted from the full GPE wavefunction at each time by projection onto the two-mode basis. (d–f) Density profiles in both rings at evolution times $\tau_0=0$, $\tau_1=T/8$, and $\tau_2=T/4$, illustrating the evolution from uniform density profiles to gray and dark solitons.}
    \label{fig:qpsvsst}
\end{figure}

We fix the inter-ring tunnelling to $t_\ell=0.2\,\hbar^2/(mL^2)$ and compare two interaction strengths, $g=0.1\,\hbar^2/(mL^2)$ and $g=1\,\hbar^2/(mL^2)$, corresponding to $t_\ell/g=2$ and $t_\ell/g=0.2$, respectively. The resulting ring currents are shown in Fig.~\ref{fig:qpsvsst}(a). For $g=0.1\,\hbar^2/(mL^2)$, the persistent current is completely transferred between the rings,  reproducing the non-interacting mode-exchange dynamics. For $g=1\,\hbar^2/(mL^2)$, by contrast, the currents change only weakly before returning close to their initial values, indicating that complete current transfer  is no longer possible.

The transition can be understood from an effective description in which each initially occupied momentum mode tunnels independently between the two rings. Projecting the coupled GPEs onto these two resonant modes, and neglecting off-resonant processes that exchange different momentum modes, yields a nonlinear two-mode model. The dynamics of either momentum component is described by its normalized population imbalance and its inter-ring relative phase,
\begin{equation}
z=
\frac{|A|^2-|B|^2}{|A|^2+|B|^2},
\qquad
\varphi=\arg B-\arg A,
\end{equation}
where $A$ and $B$ are the amplitudes of one particular momentum mode in the left and right rings. The resulting effective Hamiltonian for continuous rings is
\begin{equation}
H_{\rm 2m}(z,\varphi)=
\frac{gN_m}{L}
\left[
-\frac{z^2}{2}
-2\Lambda
\sqrt{1-z^2}\cos\varphi
\right],
\qquad
\Lambda=
\frac{t_\ell}{gN_m},
\label{eq:two-mode-continuous}
\end{equation}
where $N_m=|A|^2+|B|^2$ is the conserved total occupation of the mode. In the normalization used throughout this work, $N_m=L=1$, so that $\Lambda=t_\ell/g$. The full projection and the corresponding derivation for the continuous and discrete GPE are given in Appendix~\ref{app:twomode}.

For an initially localized mode, $z(0)=\pm1$, the effective Hamiltonian predicts complete transfer only when $
\Lambda>\frac14$.
Below this threshold, the constant-energy trajectory cannot cross $z=0$, and the mode remains self-trapped in the initial ring. The two examples shown in Fig.~\ref{fig:qpsvsst}(a) lie on opposite sides of this boundary: $t_\ell/g=2>1/4$ produces complete mode transfer, whereas $t_\ell/g=0.2<1/4$ lies in the self-trapping regime.

The phase-space structure predicted by the effective Hamiltonian is shown in Figs.~\ref{fig:qpsvsst}(b) and \ref{fig:qpsvsst}(c) for $t_\ell/g=2$ and $t_\ell/g=0.2$, respectively. Blue lines represent constant-energy trajectories of $H_{\rm 2m}(z,\varphi)$, while the dots are obtained by projecting the full GPE evolution onto the two relevant momentum modes. For $t_\ell/g=2$, the trajectories connect $z=1$ and $z=-1$, demonstrating complete transfer of each occupied momentum mode between the rings. For $t_\ell/g=0.2$, trajectories initialized near $z=\pm1$ remain confined to the same sign of $z$, confirming that the loss of current transfer originates from mode self-trapping.

The crosses labelled $\tau_0$, $\tau_1$, and $\tau_2$ mark the three times indicated by the vertical dashed lines in Fig.~\ref{fig:qpsvsst}(a): $\tau_0=0$, $\tau_1=T/8$, and $\tau_2=T/4$. The  density profiles in panels (d), (e) and (f) show the interference between the two occupied momentum modes within each ring at these three times. Initially, each ring contains a single plane wave with uniform density. As the second momentum mode enters the ring, an interference pattern develops. At $\tau=T/4$, complete mode transfer produces equal amplitudes of the two modes in each ring, generating nodes in the density profile and dark-soliton-like structures. In the self-trapped regime the amplitudes never become equal, and the density minima therefore remain finite.

Besides identifying the self-trapping threshold, the effective model predicts the interaction-induced increase of the oscillation period. For continuous rings, $T=\frac{2}{t_\ell}K\!\left(\frac{g^2N_m^2}{16t_\ell^2}\right)$
while for discrete rings the corresponding dimensionless interaction parameter and oscillation period are
$\Lambda_{\rm d}
=
\frac{2t_\ell}{UN_p}$,
and $T=\frac{2N_s}{t_\ell}K\!\left(\frac{U^2N_p^2}{64t_\ell^2}\right)$, where $K(m)$ is the complete elliptic integral of the first kind with parameter $m$.
Both expressions reduce to the non-interacting periods, $T=\pi/t_\ell$ and $T=\pi N_s/t_\ell$, respectively, while approaching the self-trapping threshold from above leads to progressively slower oscillations.

The two-mode reduction therefore establishes the regime of validity of the mode-exchange mechanism. Moderate interactions mainly slightly increase the oscillation period and generate density modulations while preserving complete current transfer. Once the self-trapping threshold is crossed, the occupied momentum modes remain localized within their original rings and coherent current exchange is suppressed.

\subsection{Limits of the phase-dominated regime}
We next analyze the robustness of the phase-dominated mechanism against stronger inter-ring tunnelling and imperfect state preparation.

The phase-dominated transport mechanism relies on the preparation of a coherent superposition of the quasi-degenerate doublet formed by the symmetric and antisymmetric combinations of localized current states. Within the proposed protocol, the quality of this preparation depends on the dimensionless interaction parameter $U/(\rho t)$. An imperfect overlap with the two-state subspace allows additional many-body eigenstates to participate in the dynamics, producing beating and a gradual loss of coherence in the current oscillations. A second source of deviations arises from increasing the inter-ring tunnelling $t_\ell$. While larger tunnel couplings accelerate the current transfer, they also move the system away from the perturbative regime, enhancing the coupling to higher-order states. In this section we investigate the robustness of the phase-dominated mechanism against these two effects by increasing $t_\ell$ for two representative interaction strengths. In particular, we quantify how the transfer period decreases beyond the perturbative limit and identify the onset of deviations from coherent two-state dynamics.

\begin{figure}[!htbp]
    \centering
    \includegraphics[width=1\linewidth]{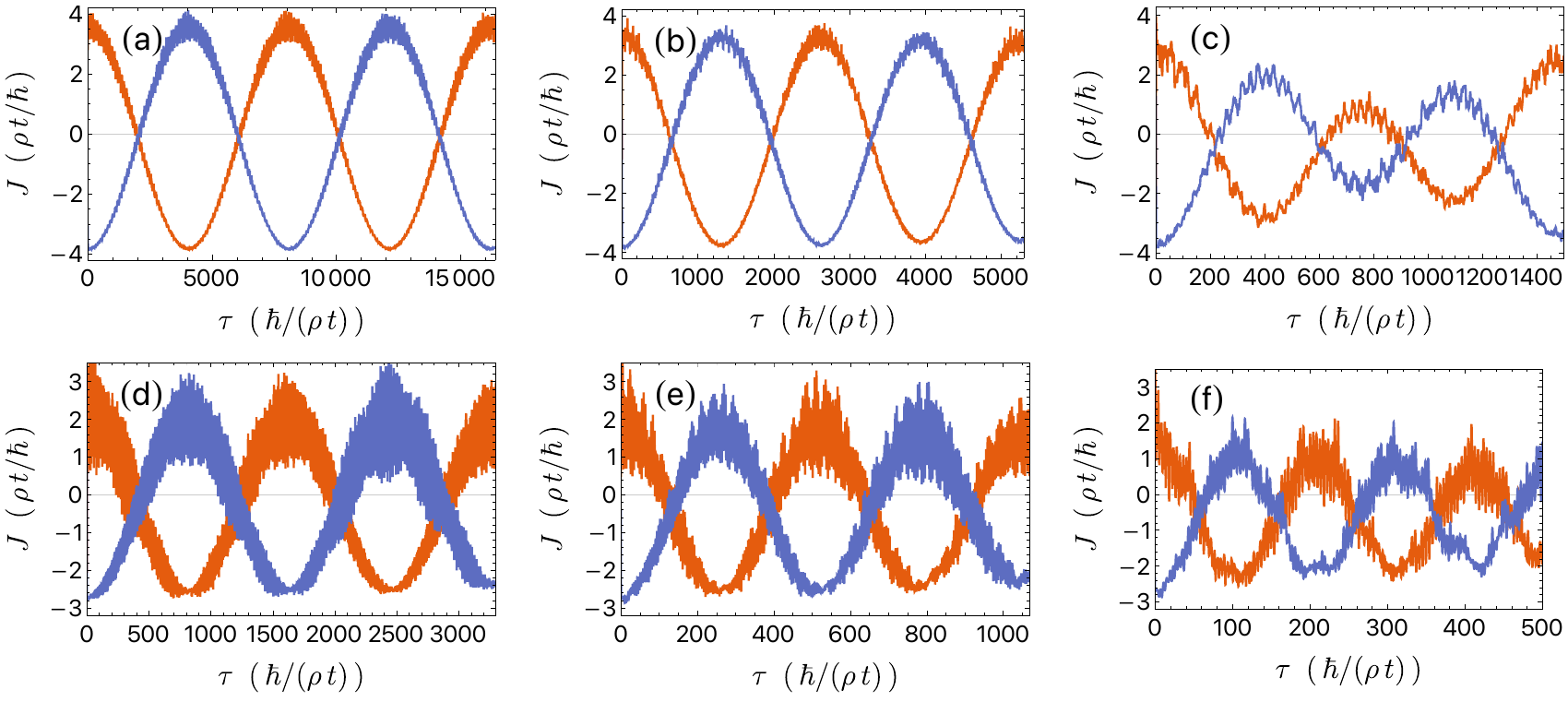}
    \caption{Persistent-current dynamics in the left (red) and right (blue) rings over two full-transfer periods. Simulations are performed using the quantum phase model for $U_{\rm QPM}=0.5$ (top row, panels (a–c)) and $U_{\rm QPM}=1.5$ (bottom row, panels (d–f)). For each interaction strength, three inter-ring tunnelling amplitudes are considered: $t_\ell/t=0.05$, $0.1$, and $0.2$, corresponding to the first, second, and third columns, respectively. Increasing $t_\ell$ accelerates the current transfer but also drives the system towards the non-perturbative regime, leading to increasingly irregular dynamics due to the excitation of additional modes.}
    \label{fig:plotsQPM}
\end{figure}

We simulate the protocol within the QPM for $U/(\rho t)=0.5$ with fluxes $\Omega_L=0.55$ and $\Omega_q=0.46$, and for $U/(\rho t)=1.5$ with $\Omega_L=1$ and $\Omega_q=0.4$. In all simulations we use a local occupation cutoff $N_{\mathrm{max}}=3$, for which all observables are well converged. For each interaction strength we compute the current dynamics for $t_\ell/t=0.05$, $0.1$, and $0.2$. The corresponding results are shown in Fig.~\ref{fig:plotsQPM}.

Despite the increasing inter-ring tunnelling, the ring-number variance remains well below one throughout the explored parameter range ($\sigma_r^2\lesssim0.7$). The dynamics therefore remains largely confined to neighboring particle-number sectors, confirming that the second-order transport mechanism discussed in the previous subsection remains valid well beyond the strict perturbative regime.

For the weaker interaction, $U/(\rho t)=0.5$, the initial state is accurately captured by the effective two-state description, resulting in nearly ideal current oscillations for $t_\ell/t=0.05$. The full oscillation period is $T=8000\,\hbar/(\rho t)$, while the current oscillates between $-4\,(\rho t)/\hbar$ and $4\,(\rho t)/\hbar$. Increasing the inter-ring tunnelling reduces the period $T$ nonlinearly, reaching $750\,\hbar/(\rho t)$ for a complete period at $t_\ell/t=0.2$, corresponding to an order-of-magnitude speedup. At this largest tunnelling, shown in Fig.~\ref{fig:plotsQPM}(c), the current is no longer fully recovered after the first oscillation, signaling the onset of the breakdown of the perturbative description.

For the stronger interaction, $U/(\rho t)=1.5$, the overlap with the effective two-state manifold is less accurate, see maps of overlaps in Appendix~\ref{app:optimization}, and other modes are already visible for $t_\ell/t=0.05$. Nevertheless, the transport remains coherent and significantly faster. The current oscillates between $-2.6\,(\rho t)/\hbar$ and $3\,(\rho t)/\hbar$ with a full period of $1600\,\hbar/(\rho t)$, already five times shorter than for $U/(\rho t)=0.5$. At $t_\ell/t=0.2$, shown in Fig.~\ref{fig:plotsQPM}(f), the full oscillation period is approximately $T=200\,\hbar/(\rho  t)$, corresponding to a first current transfer at $\tau_{\mathrm{tr}}=T/2\simeq100\,\hbar/(\rho  t)$. The eightfold reduction in period relative to $t_\ell/t=0.05$ is smaller than the factor of sixteen predicted by the perturbative scaling, indicating quantitative corrections at the largest coupling.

Since the transfer time scales inversely with the average filling $\rho $, it can be  reduced by increasing the atomic density while simultaneously increasing the interaction strength so that the ratio $U/(\rho\, t)$ remains constant. Taking $t/h=1\,\mathrm{Hz}$ and an average filling $\rho =10$ as representative experimental values, the fastest current transfer, $T/2$, presented in plot (f), corresponds to approximately $1.6\,\mathrm{s}$. This implies, given that in this case $U/(\rho t)=1.5$,  an interaction of $U/t=15$. Additional simulations with nearby values of the fluxes and interactions lead to the same qualitative conclusions and indicate that larger values of $U/(\rho t)$ are generally more robust against imperfections in the state preparation.

\section{Experimental scales and prospects}
\label{sec:expparams}

Although our analysis is independent of a specific implementation, its dimensionless parameters can be related to realistic experimental scales. A relevant benchmark is the platform of microlens-generated optical dipole traps discussed in Ref.~\cite{Sturm2017}, which explicitly considers two ring lattices connected by a tunable weak link. Its spacing, $d=1.7\,\mu\mathrm{m}$, and local control of the trapping potentials provide a suitable starting point for our geometry. The inter-ring coupling can be adjusted through the ring separation or an additional blue-detuned barrier. Below, we use the calculated Bose--Hubbard parameters of that work as order-of-magnitude estimates, without recalculating the localized orbitals for the ring configuration. We consider $N_s=3$ sites per ring; the two $^7$Li examples have $\rho=7$, corresponding to $N_p=42$ atoms in total. We quote the time to the first current transfer, $T/2$, rather than the full oscillation period.

For the mode-exchange regime, the shallow-trap estimates of Ref.~\cite{Sturm2017} imply $t/h\sim370$\,Hz for $^7$Li. Retaining its out-of-plane (light-sheet) confinement and using $U\propto a_s$, a scattering length of approximately $12a_0$ gives $U/t\sim0.02$. With $\rho=7$ and $t_\ell/t=0.2$, these parameters reproduce our mode-exchange example, for which $\tau_{\mathrm{tr}}\simeq25\hbar/t\sim11$\,ms. This estimate lies near the lower trap-depth limit considered in that reference and therefore requires a separate check of the single-band approximation.

For the phase-dominated regime, a more direct benchmark is the $^7$Li configuration in Table~I of Ref.~\cite{Sturm2017}, with $t/h=18.6$\,Hz and $U/t=10$ at $a_s\simeq100a_0$. Keeping the confinement unchanged and tuning $a_s$ to approximately $105a_0$ gives $U/t=10.5$. At $\rho=7$, this yields $U/(\rho t)=1.5$, matching our faster phase-dominated example. With the same flux protocol and $t_\ell/t=0.2$, its transfer time $\tau_{\mathrm{tr}}\simeq100\hbar/(\rho t)$ corresponds to approximately $0.12$\,s. As a slower alternative, Table~II of Ref.~\cite{Sturm2017} gives $U/t=15$ for $^{87}$Rb at an in-plane depth of approximately $k_B\times200$\,nK. Taking tunnelling on the hertz scale gives a transfer time of order $1$--$2$\,s at $\rho=10$, corresponding to $N_p=60$.

The protocol of Sec.~\ref{sec:protocol} provides one route to preparing the required initial states. Depending on the implementation, phase imprinting~\cite{phase_imprinting}, stirring~\cite{Wright2013}, or optical angular-momentum transfer~\cite{Ryu2007} could provide alternatives, although their overlap with the relevant correlated states would need to be assessed in the phase-dominated regime. For the $^7$Li phase-dominated example, the photon-scattering rate estimated in Ref.~\cite{Sturm2017}, $\Gamma_{\mathrm{sc}}\simeq0.01\,\mathrm{s}^{-1}$ per atom, corresponds to approximately $0.05$ scattering events across the sample during one transfer. This suggests that scattering from the trapping light is not a dominant limitation on this timescale, although other sources of decoherence must be assessed for a specific implementation. The required flux control and the condition that $U\rho$ remain sufficiently below the excitation gaps in all confined directions must also be established. These estimates identify plausible experimental scales without prescribing a unique implementation.

\section{Discussion and conclusions}

We have shown that macroscopic persistent currents can be coherently transferred between two weakly coupled identical rings while the average particle population of each ring remains essentially constant. Using a quench protocol that can be implemented with current ultracold-atom technology, we demonstrate that this mechanism remains operative in the many-body limit and is consistently described by the Bose--Hubbard model together with its effective Gross--Pitaevskii and Quantum Phase model limits in their respective regimes of validity.

Although the macroscopic current dynamics is similar in the two regimes considered, we find that it originates from two fundamentally different microscopic mechanisms. 
At weak interactions, circulation transfer arises mostly from independent single-particle tunnelling processes, with each occupied momentum mode behaving as an effective two-level system between the rings, defining the mode-exchange regime. At stronger interactions, the dynamics instead involves the coherent hybridization of collective many-body persistent-current states, giving rise to an effective two-state dynamics largely confined to a quasi-degenerate current doublet, which we identify as the phase-dominated regime.
These two mechanisms are distinguished by different transfer-time scalings, $T\propto t_\ell^{-1}$ and $T\propto t_\ell^{-2}$. This is also reflected by qualitatively different particle-number fluctuations, condensate fragmentation and left-right entanglement. 
In particular, mode exchange generates entanglement distributed over many Schmidt components, reflecting independent single-particle tunnelling, whereas the phase-dominated regime produces up to approximately one bit of accessible entanglement concentrated in just two dominant Schmidt components within each number sector, reflecting the coherent exchange of collective many-body current states. These distinct entanglement structures provide clear signatures of the underlying microscopic transport mechanisms.

An important distinction between the mean-field and quantum-phase models is that, although both are derived from the same Bose–Hubbard Hamiltonian, they rely on different approximations and describe different limiting regimes of the same underlying system.
The Gross–Pitaevskii description treats the condensate as a single classical field, capturing mean-field circulation dynamics but neglecting quantum correlations and tunnelling between distinct many-body current states. In contrast, the quantum phase model retains the quantum nature of the phase degrees of freedom, allowing circulation transfer through coherent tunnelling between different winding-number configurations. This distinction explains why current oscillations can occur in both descriptions, while quantum phase slips are absent from deterministic mean-field dynamics.
Increasing the interaction strength within the mean-field equations therefore does not necessarily recover the collective quantum dynamics described by the quantum phase model. In particular, extrapolating the mean-field equations to strong nonlinearities can predict current self-trapping, whereas the full Bose–Hubbard description allows quantum tunnelling between distinct many-body current configurations. 
Consequently, mean-field self-trapping is not strictly permanent in the full many-body dynamics, but coherent circulation exchange is only recovered once the phase-dominated regime is reached.

Previous mean-field studies of connected two-dimensional ring geometries have demonstrated the transfer and oscillation of quantized circulation through the spatial dynamics of the condensate density, phase and vorticity~\cite{brand2009rotational,Bland2022,chaika2026acoustic}. Here, instead, the two rings are treated as distinct quantum subsystems coupled through a localized weak link. This formulation allows us to resolve the microscopic mechanisms underlying current oscillations and characterize their many-body nature through particle-number fluctuations, condensate fragmentation, and the entanglement generated between the rings. In particular, the emergence of entanglement between distinct persistent-current configurations reveals coherent many-body dynamics that cannot be captured by a mean-field description. Our results therefore complement hydrodynamic and vortex-based approaches by providing a quantum many-body perspective on circulation transfer and the associated quantum phase-slip dynamics.

Beyond identifying these microscopic mechanisms, our results provide a broader perspective on coherent transport in atomtronic systems. In contrast to the familiar double-well problem, where the central dynamical variables are the particle imbalance and the relative condensate phase, here the transferred object is a persistent-current state. Since persistent-current states are characterized by quantized phase winding and are directly experimentally observable, they provide a natural degree of freedom for coherent manipulation and for encoding quantum information.

Several directions naturally follow from this work. Alternative state-preparation protocols, such as phase imprinting, may provide a more direct route to preparing the relevant current doublet in the phase-dominated regime and thereby improve the fidelity of the resulting oscillations. An interesting direction is to connect the present many-body description with effective orbital-angular-momentum models developed for two-dimensional lattices of coupled rings, in which the local circulation states themselves constitute the lattice degrees of freedom~\cite{polo2016geometrically,pelegri2019diamond,pelegri2019caging,pelegri2019magnetism,nicolau2023caging}. Such a comparison could establish how the effective couplings between current states emerge and change between the mode-exchange and phase-dominated regimes, and how the present mechanism generalizes from two rings to extended ring networks. 
More broadly, coherent transfer of persistent-current states could constitute a basic operation in atomtronic networks, while the effective two-state dynamics identified in the phase-dominated regime may provide a natural platform for coherent current-state qubits. A further perspective can be found in analogues to RSFQ superconducting electronics, where individual flux quanta act as discrete carriers of information in digital SQUID architectures \cite{semenov1991rsfq,febvre2010superconductive}. In an atomtronic setting, the coherent transfer of quantized circulation demonstrated here could provide a building block, with winding-number states rather than magnetic-flux quanta serving as the discrete degree of freedom. Extending this mechanism to larger ring networks could therefore open routes toward information processing and sensing architectures.

\section*{Acknowledgements}

A.P.-O. acknowledges financial support from grant PID2024-160393NB-I00, funded by MICIU/AEI/10.13039/501100011033 and by ERDF/EU.
M.B. was supported by the Los Alamos National Laboratory LDRD program under award 20230188DR.

\section*{Data availability}
The data that support the findings of this study are available upon reasonable request from the authors.

\newpage

\appendix

\section{Non--interacting Bose--Hubbard}
\label{app:bh}

To solve the non-interacting BH Hamiltonian dynamics, we introduce the momentum operators
\begin{equation}
b_{\alpha,m}=\frac{1}{\sqrt{N_s}}
\sum_{j=0}^{N_s-1}
e^{-i\frac{2\pi m j}{N_s}}a_{\alpha,j},
\qquad
a_{\alpha,j}=\frac{1}{\sqrt{N_s}}
\sum_{m=0}^{N_s-1}
e^{i\frac{2\pi m j}{N_s}}b_{\alpha,m}.
\end{equation}
The kinetic terms diagonalize as
\begin{equation}
H_\alpha=
\sum_{m=0}^{N_s-1}
\varepsilon_m(\Omega_\alpha)
\,b^\dagger_{\alpha,m}b_{\alpha,m},
\qquad
\varepsilon_m(\Omega_\alpha)
=
-2t
\cos\!\left(
\frac{2\pi}{N_s}(m-\Omega_\alpha)
\right).
\end{equation}
The link becomes
\begin{equation}
H_I=
-\frac{t_\ell}{N_s}
\sum_{m,m'}
\left(
b^\dagger_{L,m}b_{R,m'}
+
b^\dagger_{R,m'}b_{L,m}
\right).
\end{equation}
We now consider two particles in total with initial state
$|\Psi(0)\rangle
=
b^\dagger_{L,q}\,
b^\dagger_{R,0}
|0\rangle$,
corresponding to a winding-$q$ plane wave on the left ring and a constant wave on the right ring.
The current for a single particle with winding $q$ and flux $\Omega$ is $J_q=\frac{2t}{\hbar}\sin\!\left(\frac{2\pi}{N_s}(q-\Omega)\right)$.
After the quench to equal fluxes
$\Omega_L=\Omega_R\equiv\Omega$,
the relevant manifold entering the dynamics is 
\begin{align}
|A\rangle &\equiv |L_q;R_0\rangle
=
b^\dagger_{L,q}b^\dagger_{R,0}|0\rangle,\\
|B\rangle &\equiv |L_0;R_q\rangle
=
b^\dagger_{L,0}b^\dagger_{R,q}|0\rangle,\\
|C\rangle &= |L_q;L_0\rangle
=
b^\dagger_{L,q}b^\dagger_{L,0}|0\rangle,\\
|D\rangle &= |R_q;R_0\rangle
=
b^\dagger_{R,q}b^\dagger_{R,0}|0\rangle ,
\end{align}
which is degenerate with energy
$E_0=\varepsilon_q(\Omega)+\varepsilon_0(\Omega)$.
Discarding the common energy $E_0$, the link produces the effective Hamiltonian, in the basis $\{|A\rangle,|B\rangle,|C\rangle,|D\rangle\}$,
\begin{equation}
H_{\rm eff}=
-\gamma
\begin{pmatrix}
0&0&1&1\\
0&0&1&1\\
1&1&0&0\\
1&1&0&0
\end{pmatrix},
\qquad
\gamma\equiv\frac{t_\ell}{N_s}.
\end{equation}
The exact evolution is
\begin{equation}
|\Psi(\tau)\rangle
=
e^{-iE_0\tau}
\Big[
\cos^2(\gamma\tau)\,|A\rangle
-\sin^2(\gamma\tau)\,|B\rangle
+\frac{i}{2}\sin(2\gamma\tau)
\big(|C\rangle+|D\rangle\big)
\Big].
\end{equation}
The probabilities of the four configurations are therefore
\begin{align}
P_A(\tau)&=\cos^4(\gamma\tau),\\
P_B(\tau)&=\sin^4(\gamma\tau),\\
P_C(\tau)&=P_D(\tau)=\frac14\sin^2(2\gamma\tau).
\end{align}
The oscillation period is  $T=\frac{\pi}{\gamma}=\frac{\pi N_s}{t_\ell}$.

The one-boson-per-ring solution immediately generalizes to an arbitrary even
number \(N_p\) of non-interacting bosons. Defining \(M=N_p/2\), the initial state is
\begin{equation}
|\Psi(0)\rangle=
\frac{
\left(b^\dagger_{L,q}\right)^M
\left(b^\dagger_{R,0}\right)^M
}{M!}|0\rangle .
\end{equation}
Since the Hamiltonian is quadratic, each occupied orbital evolves independently according to the single-particle solution,
\begin{align}
b^\dagger_{L,q}
&\rightarrow
c\,b^\dagger_{L,q}
-is\,b^\dagger_{R,q},
\\
b^\dagger_{R,0}
&\rightarrow
c\,b^\dagger_{R,0}
-is\,b^\dagger_{L,0},
\end{align}
with \(c=\cos(\gamma\tau)\) and \(s=\sin(\gamma\tau)\). The many-body state therefore reads
\begin{equation}
|\Psi(\tau)\rangle=
\frac{
\left(c\,b^\dagger_{L,q}-is\,b^\dagger_{R,q}\right)^M
\left(c\,b^\dagger_{R,0}-is\,b^\dagger_{L,0}\right)^M
}{M!}|0\rangle ,
\end{equation}
from which all observables follow from independent binomial statistics. Their exact analytical expressions are summarized in Table~\ref{tab:non_int_bh}.

\begin{table}[th]
\centering
\begin{tabular}{l c}
Observable & Non-interacting ($U=0$) \\
\hline
Period &
$\displaystyle T=\frac{\pi N_s}{t_\ell}$ \\[1ex]

Ring population &
$\displaystyle
\langle N_{L,R}\rangle=\frac{N_p}{2}
$ \\[1ex]

Ring-number variance &
$\displaystyle
\sigma_{N_{L,R}}^2=
\frac{N_p}{4}
\sin^2(2\gamma\tau)
$ \\[1ex]

Site-number variance &
 $\displaystyle
\sigma_n^2=\rho
\left(
1-\frac{1}{N_s}
\right)
+
\frac{\rho}{2}
\left(
\rho+\frac{1}{N_s}
\right)
\sin^2(2\gamma\tau)
$ \\[1ex]

Current (left) &
$\displaystyle
\langle J_L\rangle=M\,J_q\cos^2(\gamma\tau)+M\,J_0\sin^2(\gamma\tau)
$ \\[1ex]

Current (right) &
$\displaystyle
\langle J_R\rangle=M\,J_q\sin^2(\gamma\tau)+M\,J_0\cos^2(\gamma\tau)
$ \\[1ex]

Ring condensate fraction &
$\displaystyle
f_c^{\rm ring}
=
\max(\cos^2,\sin^2)
$ \\[1ex]

Global condensate fraction &
$\displaystyle
f_c^{\rm tot}=\frac12
$
\end{tabular}
\caption{Exact many-body solution for non-interacting bosons.}
\label{tab:non_int_bh}
\end{table}

\section{Two-mode Hamiltonian}
\label{app:twomode}

We assume the coupled continuous GPEs model and a two-mode ansatz,
\begin{align}
\psi_L(x,t) &= \frac{1}{\sqrt{L}}\Big(A_1(t)e^{ik_1 x}+A_2(t)e^{ik_2 x}\Big),\label{eq:ansatz_L}\\
\psi_R(x,t) &= \frac{1}{\sqrt{L}}\Big(B_1(t)e^{ik_1 x}+B_2(t)e^{ik_2 x}\Big),\label{eq:ansatz_R}
\end{align}
with $k_1=\frac{2\pi}{L}q_1$, $k_2=\frac{2\pi}{L}q_2$. Initially we set $A_1(0)=B_2(0)=1$, $A_2(0)=B_1(0)=0$.
Using polar form,
\begin{equation}
A_1=r_{1a}e^{i\alpha_1},\qquad
A_2=r_{2a}e^{i\alpha_2},\qquad
B_1=r_{1b}e^{i\beta_1},\qquad
B_2=r_{2b}e^{i\beta_2},
\label{eq:polar_def}
\end{equation}

Substituting this ansatz in the coupled GPEs, integrating out fast rotating terms, with frequencies $|\beta_2-\alpha_1|,~|\alpha_2-\beta_1|\gg |\alpha_1-\beta_1|,~|\alpha_2-\beta_2|$, we can focus on mode 1, and work out the equation for the population imbalance (of a mode, not of particles). Defining,
\begin{equation}
\varphi_1 \equiv \beta_1-\alpha_1,
\qquad
z_1=\frac{|A_1|^2-|B_1|^2}{|A_1|^2+|B_1|^2}.
\label{eq:defs_z1_swap}
\end{equation}
The $(q_1)$ dynamics closes on $(z_1,\varphi_1)$ as
\begin{align}
\dot z_1&=
-\frac{2t_\ell}{L}\,\sqrt{1-z_1^2}\,\sin\varphi_1,\\
\dot\varphi_1&=
-\frac{g}{L}N_1 z_1
+\frac{2t_\ell}{L}\,\frac{z_1}{\sqrt{1-z_1^2}}\,\cos\varphi_1,
\end{align}
with $N_1=|A_1|^2+|B_1|^2$ constant, and in our simulations we use the normalization $N_1=1$ and the perimeter of the ring is $L=1$. These two variables have a Hamiltonian
\begin{equation}
H(z_1,\varphi_1)
=
\frac{g\,N_1}{L}\left[
-\frac{1}{2} z_1^2
-2\frac{t_\ell}{g\,N_1}\sqrt{1-z_1^2}\,\cos\varphi_1\right].
\label{eq:H_final}
\end{equation}
One can check it satisfies
\begin{equation}
\dot z_1=-\frac{\partial H}{\partial \varphi_1},\quad
\dot\varphi_1=\frac{\partial H}{\partial z_1}.
\label{eq:canonical}
\end{equation}

\subsection{Corresponding 2-mode reduction for coupled dGPEs}

The derivation for the coupled discrete GPEs is analogous to the case for continuous rings. 
Now, the allowed lattice quasimomenta are
\begin{equation}
q_m=\frac{2\pi m}{N_s},
\qquad m=0,\ldots,N_s-1.
\end{equation}
We retain the two quasimomenta, $q_1$ and $q_2$, and write the ansatz as
\begin{align}
\phi_{L,j}(t)
&=
\frac{1}{\sqrt{N_s}}
\left[
A_1(t)e^{iq_1j}
+
A_2(t)e^{iq_2j}
\right],
\label{eq:dGPE_ansatz_left}
\\
\phi_{R,j}(t)
&=
\frac{1}{\sqrt{N_s}}
\left[
B_1(t)e^{iq_1j}
+
B_2(t)e^{iq_2j}
\right].
\label{eq:dGPE_ansatz_right}
\end{align}
Now the effective two-mode Hamiltonian is:
\begin{equation}
H(z_1,\varphi_1)
=\frac{U\,N_p}{2N_s}\left[
-\frac{1}{2}z_1^2
-2\,\frac{2t_\ell}{U\,N_p}\sqrt{1-z_1^2}\cos\varphi_1\right].
\end{equation}

\section{From the Bose--Hubbard model to the quantum phase model}
\label{app:qpm}

The quantum phase model (QPM) can be understood as the large-filling limit of the Bose--Hubbard model (BHM), where the average occupation per site is large while number fluctuations remain small. This mapping provides an intuitive justification for comparing both models in the parameter regime considered in the main text, where the filling is sufficiently large and interactions are strong. Alternative derivations and validations of the QPM can be found in the literature; here we simply present the standard large-filling limit of the BHM.

The one-dimensional Bose--Hubbard Hamiltonian on a ring is

\begin{equation}
H=
\frac{U}{2}\sum_{i=0}^{N_s-1}\hat n_i(\hat n_i-1)
-t\sum_{i=0}^{N_s-1}\left(\hat a_i^\dagger\hat a_{i+1}+\hat a_{i+1}^\dagger\hat a_i\right).
\end{equation}

We assume a large average occupation per site $\rho $ and write
\begin{equation}
\hat n_i=\rho +\delta\hat n_i,
\qquad
|\delta\hat n_i|\ll\rho,
\qquad
\sum_{i=0}^{N_s-1}\delta\hat n_i=0.
\end{equation}
The interaction term becomes
\begin{align}
\frac{U}{2}\sum_{i=0}^{N_s-1}\hat n_i(\hat n_i-1)
=&\,
\frac{U}{2}
\sum_{i=0}^{N_s-1}
(\rho+\delta\hat n_i)
(\rho+\delta\hat n_i-1)
\nonumber\\
=&\,
\frac{U}{2}N_s(\rho^2-\rho)
+\frac{U}{2}\sum_{i=0}^{N_s-1}(\delta\hat n_i)^2,
\end{align}
where the linear contribution vanishes because the total particle number is conserved.

For the hopping term,
\begin{align}
\hat a_i^\dagger\hat a_j
&
\left|
\rho+\delta n_i,\,
\rho+\delta n_j
\right\rangle
\nonumber\\
&=
\sqrt{(\rho+\delta n_i+1)(\rho+\delta n_j)}
\left|
\rho+\delta n_i+1,\,
\rho+\delta n_j-1
\right\rangle
\nonumber\\
&\simeq
\left(
\rho+\mathcal O\!\left(\frac{\delta n}{\rho}\right)
\right)
\left|
\rho+\delta n_i+1,\,
\rho+\delta n_j-1
\right\rangle,
\end{align}
so that, to leading order in $\delta n/\rho$, the hopping matrix element is proportional to the average occupation $\rho$. Using the standard phase approximation, $\hat a_j\simeq\sqrt{\rho}\,e^{i\hat\phi_j}$,
with
\begin{equation}
e^{\pm i\hat\phi_j}
|\delta n_j\rangle
=
|\delta n_j\mp1\rangle,
\end{equation}
the hopping term becomes
\begin{equation}
-t\rho
\sum_{i=0}^{N_s-1}
\left[
e^{i(\hat\phi_i-\hat\phi_{i+1})}+e^{i(\hat\phi_{i+1}-\hat\phi_i)}
\right].
\end{equation}
Discarding the constant interaction energy, the Hamiltonian reduces to
\begin{equation}
H=
\frac{U}{2}\sum_{i=0}^{N_s-1}(\delta\hat n_i)^2
-
2t_{\rm QPM}
\sum_{i=0}^{N_s-1}
\cos(\hat\phi_{i+1}-\hat\phi_i),
\end{equation}
where $t_{\rm QPM}=\rho\,t$.
A Peierls phase is incorporated exactly as in the Bose--Hubbard model,
\begin{equation}
H=
\frac{U}{2}\sum_{i=0}^{N_s-1}(\delta\hat n_i)^2
-2t_{\rm QPM}\sum_{i=0}^{N_s-1}\cos(\hat\phi_{i+1}-\hat\phi_i-2\pi\Omega/N_s)
\end{equation}

\section{Optimization of the preparation protocol in the phase-dominated regime}
\label{app:optimization}

In the phase-dominated regime, coherent current transfer requires the initial product state to have a large overlap with the nearly degenerate symmetric and antisymmetric eigenstates of the post-quench Hamiltonian. Since the right-ring flux is unchanged by the quench, $\Omega_R=\Omega_q$, its initial ground state is already an eigenstate of the post-quench single-ring Hamiltonian. The only optimization therefore concerns the preparation flux $\Omega_L$ of the left ring.

For each pair of fluxes $(\Omega_L,\Omega_q)$, we diagonalize the single-ring QPM Hamiltonian. We compute the ground state for the preparation flux,
$|\psi_L^{(0)}\rangle$,
and all eigenstates of the Hamiltonian at the post-quench flux, $|\psi_q^{(n)}\rangle$.
The relevant preparation fidelity is defined as
\[
F(\Omega_L,\Omega_q)
=
\max_n
\left|
\langle
\psi_q^{(n)}
|
\psi_L^{(0)}
\rangle
\right|^2,
\]
where the maximization selects the post-quench eigenstate with the largest overlap with the prepared ground state.

\begin{figure}[th] 
    \centering
    \includegraphics[width=1\linewidth]{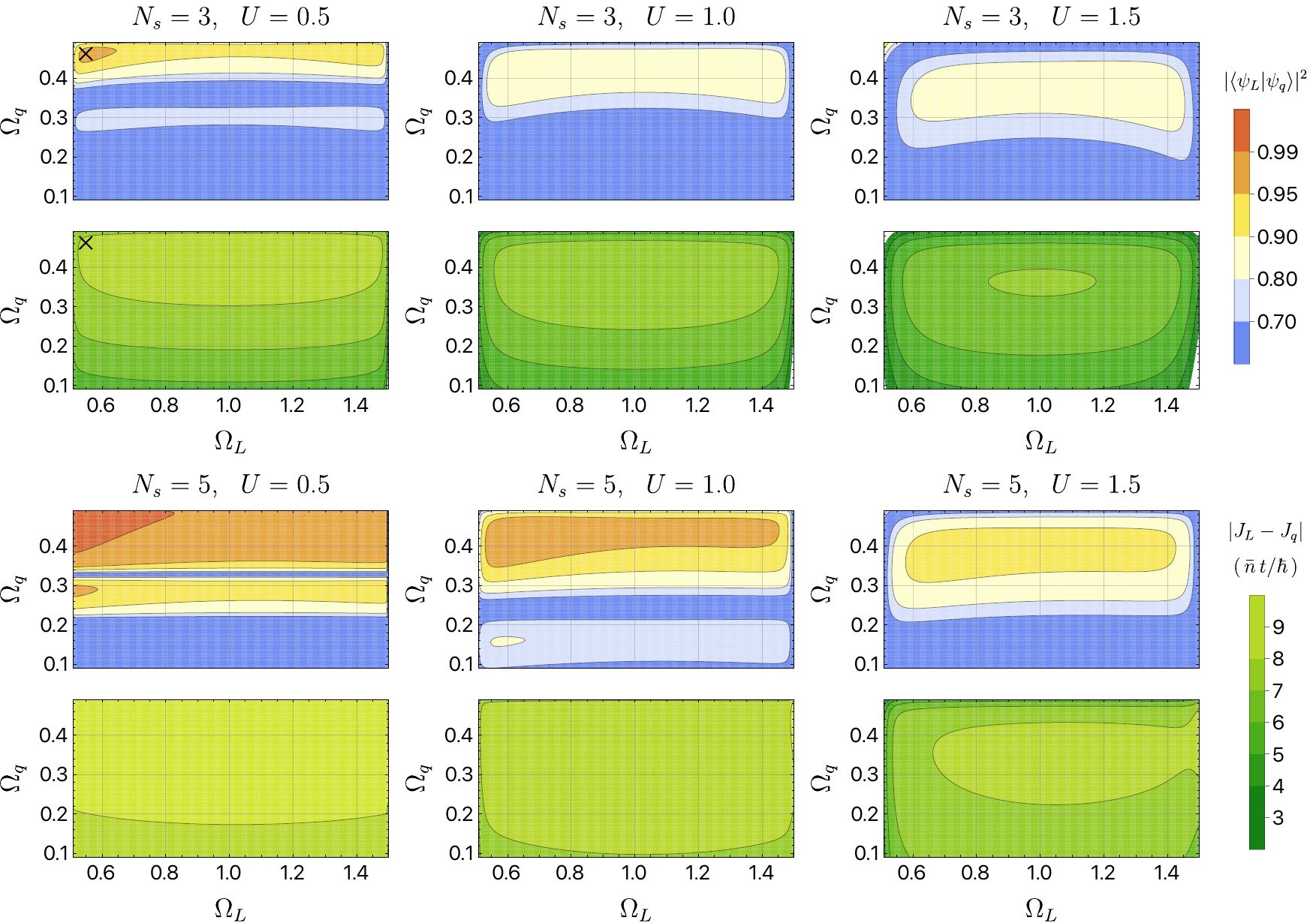}
    \caption{Preparation fidelity $F(\Omega_L,\Omega_q)$ (first and third rows) and current contrast $J_L-J_q$ in units of $\rho t/\hbar$ (second and fourth rows) for $N_s=3$ (top) and $N_s=5$ (bottom) and $U_{\rm QPM}=0.5,1.0,1.5$. The cross marks the parameters used in Fig.~\ref{fig:universal_current}(b).}
    \label{fig:overlaps}
\end{figure}

The resulting fidelities are shown in the first and third rows of Fig.~\ref{fig:overlaps} for two interaction strengths, $U_{\rm QPM}=0.5$ and $1.5$, and two lattice sizes, $N_s=3$ and $5$. To ensure that the selected post-quench state carries a current significantly different from the ground state at $\Omega_q$, we also compute $|J_L-J_q|$,
where $J_q$ is the current of the ground state at $\Omega_q$ and $J_L=\langle\psi_L^{(0)}|\hat J(\Omega_q)|\psi_L^{(0)}\rangle$ is the current of the prepared state immediately after the quench, evaluated with the post-quench flux. To avoid the degeneracy between clockwise and counterclockwise current states, we restrict the search to $\Omega_L>0.5$ and $0.1<\Omega_q<0.5$.

For both lattice sizes, the largest overlaps are obtained for the weaker interaction, $U_{\rm QPM}=0.5$, with $\Omega_L\gtrsim0.5$ and $\Omega_q\lesssim0.5$. The maximum fidelity exceeds $0.95$ for $N_s=3$ and $0.99$ for $N_s=5$. Increasing the interaction to $U_{\rm QPM}=1.5$ reduces the overlap, although values around $0.8$ ($N_s=3$) and $0.93$ ($N_s=5$) are still obtained.

The corresponding current difference remains approximately independent of $\Omega_L$ and reaches
$|J_L-J_q|
\simeq
8\,\rho t/\hbar$ ($N_s=3$) and $9.5\,\rho t/\hbar$ ($N_s=5$)
in the region of optimal overlap. Although the current-difference map is quantitatively different for $N_s=3$ and $5$, the optimal values are very similar. Overall, increasing the ring size improves the state-preparation fidelity while preserving a comparable current contrast, explaining the cleaner current oscillations observed in the full dynamical simulations for $N_s=5$ (not shown).

The cross in the upper-left panels of Fig.~\ref{fig:overlaps} indicates the parameters used for the Bose--Hubbard and QPM simulations shown in Fig.~\ref{fig:universal_current}(b).

\section{Bose--Hubbard numerical results}
\label{app:sims}

Table~\ref{tab:appendixResults} summarizes the Bose--Hubbard simulations for different particle numbers in the two transport regimes discussed in the main text. These results support the scaling relations and characteristic signatures summarized in Table~\ref{tab:regimesDiff}. In particular, the mode-exchange regime exhibits extensive ring-number fluctuations and a transfer period that is nearly independent of particle number, whereas the phase-dominated regime is characterized by strongly suppressed number fluctuations, an almost perfectly two-dimensional Schmidt subspace ($W_2\simeq1$), and a transfer period that decreases approximately as $T\propto N_p^{-1}$, as predicted by the QPM.

\begin{table}[th]
\centering
\begin{tabular}{c|cccc|cccc}
 & \multicolumn{4}{c|}{Mode-exchange (mean field)}
 & \multicolumn{4}{c}{Phase-dominated} \\
\hline
$N_p$
& 30 & 36 & 42 & 48
& 30 & 36 & 42 & 48 \\
\hline
$U\,\rho/t$, $U/(\rho\,t)$
& 0.10 & 0.12 & 0.14 & 0.16
& 0.50 & 0.50 & 0.50 & 0.50 \\
$T\,(\hbar/t)$
& 48.8 & 52.0 & 53.7 & 54.1
& 758 & 554 & 462 & 407\\
$\overline{\sigma_r^2}/(N_p/2)$
& 0.248 & 0.261 & 0.265 & 0.266
& 0.043 & 0.028 & 0.022 & 0.017\\
$\overline{\sigma_s^2}$
& 11.0 & 15.1 & 19.7 & 25.2
& 0.83 & 0.81 & 0.81 & 0.81\\
$W_2$
& 0.483 & 0.423 & 0.379 & 0.328
& 0.9995 & 0.9997 & 0.9996 & 0.9989\\
\end{tabular}
\caption{
Numerical Bose--Hubbard results supporting the characteristic properties summarized in Table~\ref{tab:regimesDiff}. Rows: interaction parameter ($U\rho/t$ for mode exchange, $U/(\rho t)$ for the phase-dominated regime), current-transfer period, ring- and site-number fluctuations, and two-Schmidt weight. Overbars denote averages over one complete current-transfer period. The quantity $W_2$ is computed within the dominant number sector.
}
\label{tab:appendixResults}
\end{table}

\bibliographystyle{unsrt}

\bibliography{references.bib}

\end{document}